\documentclass{main}

\usepackage{amsmath}

\def\be{\begin{equation}}
\def\ee{\end{equation}}
\def\bea{\begin{eqnarray}}
\def\eea{\end{eqnarray}}

\newcommand{\Photo}{\vskip .2 in \includegraphics[height=35mm]{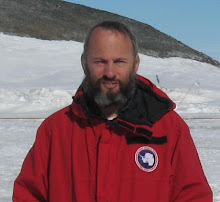}\vskip .15 in}

\begin{document}

\title{\Large Quantum mechanical aspects of coherent photoproduction: \\the limits of coherence, and multiple vector mesons }

\author{Spencer R. Klein}

\address{Nuclear Science Division, Lawrence Berkeley National Laboratory\\Berkeley CA 94720 USA}

\maketitle\abstracts{Quantum mechanics is central to coherent photoproduction in ultra-peripheral collisions  (UPCs).  This writeup will discuss some surprising aspects of UPCs that stem from these quantum mechanical roots.  The Good-Walker (GW) paradigm, which connects coherent photoproduction with the target nucleus remaining in its ground state.  This contrasts with a semi-classical picture, where coherence depends on the positions of the individual nucleons and the momentum transfer.  Unlike the GW approach, the semiclassical picture is consistent with the observed data on coherent photoproduction with nuclear breakup, and with coherent photoproduction in peripheral collisions.  The semiclassical approach allows for a wider variety of coherent UPC reactions, such as coherent photoproduction of charged mesons, including some non $q\overline q$ exotica.  
Quantum mechanics is also key to the coherent photoproduction of multiple vector mesons by the interactions of a single ion pair.  The vector mesons share a common impact parameter, and so can exhibit richer interference patterns than single mesons.  At forward rapidities, the cross sections to produce multiple identical vector mesons are enhanced due to superradiance.  With enough statistics, multi-meson events may provide an opportunity to observe stimulated decays.}


\section{Introduction}

Ultra-peripheral collisions (UPCs) are interactions where relativistic nuclei interact at impact parameters $b$ greater than twice the nuclear radius ($R_A$), where there are no hadronic interactions \cite{Bertulani:2005ru,Baltz:2007kq,Contreras:2015dqa,Klein:2020fmr}.   Instead the ions can interact electromagnetically, via either two-photon processes or photonuclear interactions.  UPCs at CERN's Large Hadron Collider (LHC) are the energy frontier for both of these classes of events.  Two-photon production of lepton pairs and photon pairs (light-by-light scattering) \cite{dEnterria:2013zqi} have been intensively studied, along with searches for beyond-standard-model physics \cite{Bruce:2018yzs,Klein:2020fmr}.  Photonuclear interactions are an important probe of nuclear structure at low Bjorken$-x$, having been studied via photon-gluon fusion to produce dijets, and via photon-Pomeron interactions which produce vector mesons.  

Exclusive production of vector mesons is of particular interest, both theoretically and experimentally.  Experimentally, the final states are usually simple - two oppositely charged particles -  which greatly simplifies detection.   Vector meson production involves the exchange of a colorless object. At leading order, at least two gluons must be exchanged, but recent next-to-leading order calculations of $J/\psi$ photoproduction have shown that there are significant cancellations involving the gluonic contributions to the cross sections, so the quark distributions play a large role \cite{Eskola:2022vpi}.   The same calculation also found that the cross section was very sensitive to the assumed pQCD scale.  

Despite these difficulties, much interest remains in using coherent vector meson production to probe nuclei because it provides access to information on the transverse positions of partons, and of their fluctuations, via the Good-Walker paradigm which links coherent production with the average nuclear configuration and incoherent production to fluctuations in the nuclear configuration, including the presence of gluonic hot spots \cite{Klein:2019qfb}.  $d\sigma/dt$ for coherent production can be Fourier transformed to find the transverse distribution of targets (gluons) in the target.   

\section{Coherent and incoherent photoproduction}

\subsection{The Good-Walker paradigm}

The total cross section for vector meson photoproduction may be written \cite{Good:1960ba}
\begin{equation}
\frac{d\sigma_{tot}}{dt} = \frac{1}{16\pi} \langle |A(K,\Omega)|^2\rangle
\label{eq:sigmatot}
\end{equation}
where $A(K,\Omega)$ is the amplitude for producing a vector meson, where $K$ represents the kinematic factors (vector meson momentum, momentum transfers, etc.) and $\Omega$ represents the configuration of the target, including the positions of the target nucleons and the number and positions of the partons within those nucleons.   The amplitudes are squared, and then averaged over all of the possible $\Omega$.  In contrast, for coherent production, the amplitudes are added (averaged) and then squared: 
\begin{equation}
\frac{d\sigma_{coh}}{dt} = \frac{1}{16\pi} | \langle A(K,\Omega)\rangle|^2.
\label{eq:sigmacoh}
\end{equation}
Mietenlin and Pumplin pointed out \cite{Miettinen:1978jb} that the difference between these is the incoherent cross section, which is sensitive to event-by-event fluctuations in the cross section:
\begin{equation}
\frac{d\sigma_{incoh}}{dt} = \frac{1}{16\pi} \big( \langle |A(K,\Omega)|^2\rangle - | \langle A(K,\Omega)\rangle|^2\big).
\label{eq:sigmaincoh}
\end{equation}
This is the sum of squares minus the square of sums, so is directly sensitive to event-by-event fluctuations in the cross sections.  

In UPCs, $|t|\approx {\rm pair}\ p_T^2$ (neglecting the generally small $t_z$).  Since $p_T$ and impact parameter ($b$) are conjugate variables, one can perform a 2-dimensional Fourier-Bessel  transform of $d\sigma/dp_T$ to get $F(b)$, the amplitude for having an interaction at an impact parameter $b$ \cite{Munier:2001nr,STAR:2017enh}:
\begin{equation}
F(b) \propto \int_0^\infty p_T dp_T J_0(bp_T)\sqrt{\frac{d\sigma}{dt}}\ ^*
\label{eq:Fourier}
\end{equation}
where $J_0(x)$ is a Bessel function.  Several caveats are associated with this transformation \cite{Klein:2021mgd}.  The asterisk is because is necessary to flip with sign of the square root when integrating across each diffractive minimum.  This is to properly track the phase of the amplitude across these zeros.   This is theoretically straightforward, but experimentally it can be difficult to pinpoint  these zeros accurately enough.   Second, the integral goes from zero to infinity, but real data has a maximum $p_T$. The presence of a $p_T$ cutoff is equivalent to applying a square windowing function, and so introduces artifacts.  

The STAR collaboration investigated this in a high-statistics study of $\rho^0$ plus direct $\pi^+\pi^-$ production\cite{STAR:2017enh}.  They found that varying the maximum $p_T$ cutoff affected $F(b)$ at small $b$, but did not change the apparent size of the target.   In UPCs, the measured vector meson $p_T$ includes contributions from the experimental resolution and from the photon $p_T$, although these can be removed by deconvolution \cite{ALICE:2021tyx}.  

\subsection{Problems with the Good-Walker approach}

Unfortunately, there is a significant problem with the overall Good-Walker approach \cite{Klein:2023zlf}.  Equation \ref{eq:sigmacoh}, identifies coherent production with the nucleus remaining unexcited, in the ground state.  However, there is another definition of coherence - the addition of amplitudes in-phase, leading to a peak at small $p_T$ ($p_T < \hbar/R_A$).  This low $p_T$ coherent peak has been observed in different types of reactions where the nuclear  targets do not remain in the ground state. 

\begin{figure}[!tbp]
    \centering
    \includegraphics[width=0.4\textwidth]{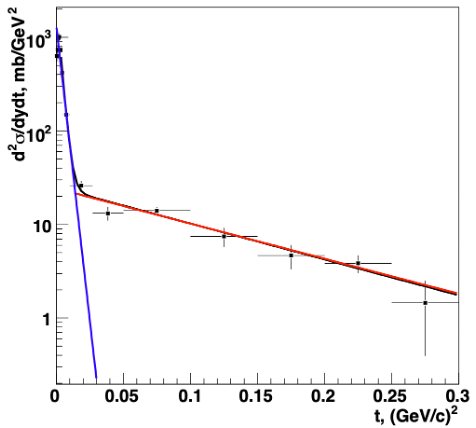} 
    \includegraphics[width=0.4\textwidth]{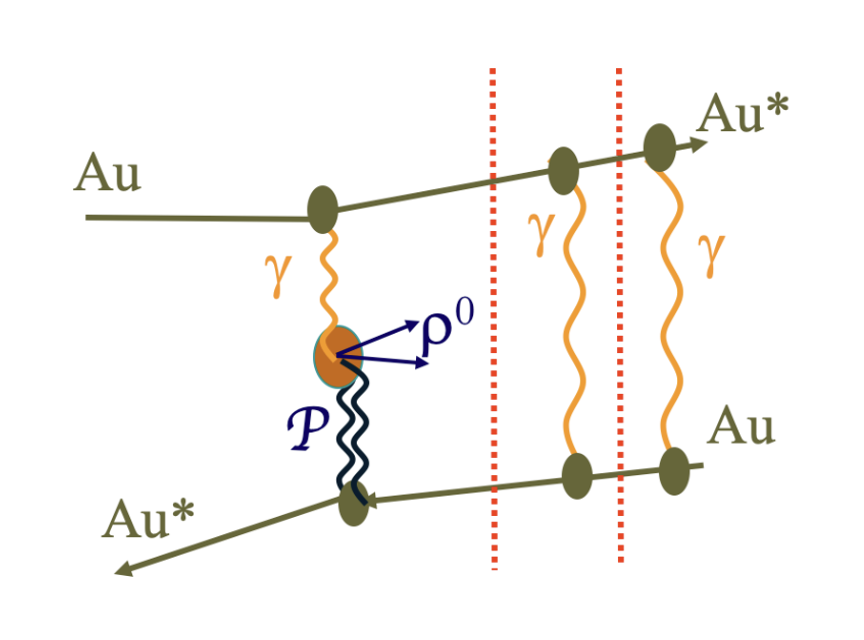} 
    \caption{(left) $d\sigma/dt$ for $\rho$ photoproduction as measured by the STAR Collaboration.    The data is well fit by two exponentials, corresponding to coherent coupling to the entire nucleus (blue) and to individual nucleons (red).   This data required that both of the nuclei be excited, as evidenced by the observation of breakup neutrons in both forward zero degree calorimeters, but a coherent peak is still clearly visible. From Ref. \protect\cite{STAR:2007elq}.  
(right) Schematic diagram showing how mutual Coulomb dissociation factorizes (decouples) from $\rho$ photoproduction; the processes separated by the vertical red dotted lines share only a common impact parameter.}
\label{fig:coherent}
\end{figure}

One such class is vector meson photoproduction accompanied by mutual excitation of both nuclei.  This has been seen by both STAR \cite{STAR:2002caw,STAR:2007elq} and ALICE \cite{ALICE:2020ugp} collaboration.  For STAR, the mutual breakup requirement is intrinsic to most of their UPC triggers.  The effect of mutual breakup is calculable by assuming factorization for multi-photon exchange, whereby one photon produces the vector meson, and two additional photons each excite one of the nuclei \cite{Baltz:2002pp}, as is shown in Fig. \ref{fig:coherent} (right).    The photons are emitted independently, save for their common impact parameter \cite{Baur:2003ar}.  Factorization does a good job of predicting the cross sections for coherent production of vector mesons accompanied by neutron emission \cite{Baltz:2002pp}.   

However, it does not fit into the Good-Walker approach, where the overall reaction is under scrutiny.   Further, in some $p_T$ range, incoherent photoproduction of vector mesons (one-photon exchange) and coherent photoproduction of vector mesons accompanied by the exchange of an additional photon which excites one of the nuclei will both contribute.  These two channels can even interfere with each other.   

One posited way out of this paradox uses time scales.  Vector meson production occurs on time scales that are short compared with the nuclear excitation, so the vector meson is effectively decoupled from the nuclear breakup.  However, the most common excitation, to a Giant Dipole Resonance (GDR) requires an energy of 10-20 MeV in the target rest frame.  But, for a high-energy photon converting to a vector meson, the required momentum transfer from the target corresponds to considerably less energy, and therefore a longer time scale.    

It is worth noting that the idea of exclusive production is slightly simplistic, in that it neglects other reactions that are likely to occur.  For two ions with a relatively small $b$ (but still with $b>2R_A$), the probability of two-photon production of $e^+e^-$ pairs is large - for collisions with lead nuclei at the LHC, the average number of pairs produced when $b\approx 2R_A$ is 3-4 \cite{Baur:2007zz}.  These pairs are generally unobservable, but they are still present.  And, although the cross section for bremsstrahlung emission by the ion is small, it is also infrared divergent, so some soft radiation is always expected.   So, the idea of completely exclusive interactions is an unattainable theoretical ideal. 

The second problematic process is vector meson production in peripheral heavy-ion collisions, where the low $p_T$ vector meson is accompanied by hadronic interactions which can produce hundreds of particles.   Coherent (again defined as exhibiting a peak at low $p_T$, consistent with the in-phase addition of amplitudes) photoproduction of $J/\psi$ has been observed in gold-gold collisions by STAR \cite{STAR:2019yox}, and in lead-lead collisions by ALICE \cite{ALICE:2022zso} and LHCb \cite{LHCb:2021hoq}.  The cross sections appear consistent with theoretical calculations that largely decouple the photoproduction from the hadronic interactions \cite{Zha:2017jch,Klusek-Gawenda:2015hja}, although the precision of the data is still limited, and it is not possible to conclusively determine the size of the coherent target.   It is not currently possible to reconcile this data with the underpinnings of the Good-Walker approach.

\subsection{A semi-classical approach}

It is possible, however, to explain all of this data in a semi-classical approach, where the cross section for vector meson production on a nucleus consisting of $i$ nucleons at positions $\vec{x}_i$ is  \cite{Klein:2023zlf}
\begin{equation} 
\sigma_{coh} = \Sigma_i \big|A_i \exp(i\vec{t}\cdot\vec{x}_i)\big|^2
\label{eq:semiclassical}
\end{equation}
Here $A_i$ is the interaction amplitude for each nucleon, and $\vec{t}$ is the 3-momentum transfer from the target.   This approach is insensitive to what happens to the target, but it still reproduces the observed p$p_T$ spectra for coherent and incoherent photoproduction.   When $|\vec{t}|$ is small, the exponential is close to one, and the cross section is the square of the sum of the amplitudes; this leaves no room for incoherent production at small $|t|$.  At large $|t|$, the exponential takes on random phases, and the cross section scales as the sum of the square of the amplitudes.  These regimes are visible in the STAR $\rho^0$ photoproduction data \cite{STAR:2007elq} shown in Fig. \ref{fig:coherent} (left).  There, the incoherent cross section still decreases slowly with increasing $p_T$ due to the form factor of the individual nucleons.  One drawback of Eq. \ref{eq:semiclassical} is that it does not track the outgoing nucleus, so one cannot separate coherent and incoherent production, except via the different $p_T$ scales - an imperfect classification.

Although the semi-classical approach makes fairly similar predictions as Good-Walker for coherent production, the interpretations for incoherent production are totally different.  In Good-Walker, it depends on fluctuations, while in the semi-classical approach, it depends only on the static properties of the target. 

\subsection{Gold vs. lead: similar, or not?}

Another issue with the Good-Walker approach appears at low $|t|$  \cite{Klein:2023zlf}.  It relates the incoherent cross section to event-by-event  fluctuations in nucleon positions and subnucleonic (parton level) fluctuations.   From these points of view, heavy nuclei are expected to be similar.   For example, the density distributions of gold-197 and lead-208  are both well described by a Woods-Saxon distribution and they have a similar partonic structure, with similar nuclear shadowing.   However, in the nuclear shell model, their structures are very different, and this their incoherent interactions must be different at low $|t|$.  

Lead-208 is doubly magic, with a lowest excited state at an energy of $E_{\rm exc.} = 2.6$ MeV.  No excitation ( is possible at lower energy transfers, so there can be no incoherent interactions.  If one assumes that the Pomeron $|t|$ is transferred to a single nucleon (as indicated by the STAR data on incoherent $\rho^0$ plus direct $\pi^+\pi^-$ photoproduction, taken at larger $|t|$ \cite{STAR:2017enh}), then the energy threshold corresponds to a minimum momentum transfer of $p=\sqrt{2m_p E_{\rm exc.}}= 71$ MeV/c.  In contrast, gold-197 has a lowest excited state at an energy of 77 keV, corresponding to a minimum $p$ of only 12 MeV/c.   This indicates that incoherent photoproduction in lead is impossible for $p_T < 71$ MeV/c, while in gold the minimum is 12 MeV/c.  Although Fermi motion and other factors may smooth out these cutoffs, but there should be large differences in incoherent production at small $|t|$. 

\subsection{Additional possibilities from the semiclassical formulation}

The semi-classical formulation allows for a wider range of coherent reactions than Good-Walker.   It allows for the coherent exchange of Reggeons that include non-zero quantum numbers such as spin, or even electrical  charge.  This permits reactions such as
\begin{eqnarray}
\gamma + A \rightarrow a_2^+(1320) + (A-1) \\
\gamma + A \rightarrow a_2^-(1320) + (A+1).
\end{eqnarray}
The first reaction changes a proton into a neutron, while the other converts a nucleon in the other direction.    In the semiclassical approach, the first reaction could be coherent over the protons in the target while the second could be coherent over the neutrons.  This should lead to a well-determined production cross section ratio, based on the cross sections on protons and on the nuclear composition of the target.   By measuring $d\sigma/dt$ for the two reactions, it would also be possible to determine the proton and neutron radii of the target nucleus, with a very small relative systematic error, allowing for a reliable measurement of the thickness of any neutron skin.  

\begin{figure}[!tbp]
    \centering
    \includegraphics[width=0.48\textwidth]{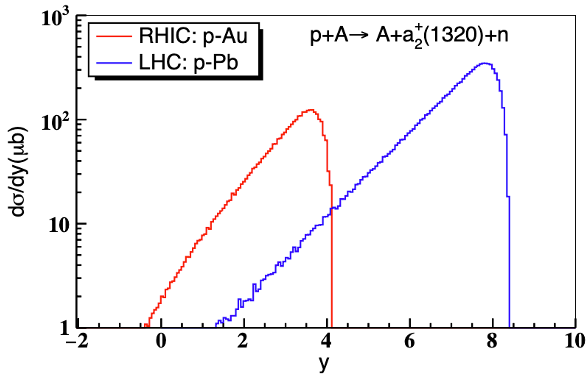} 
    \caption{Predicted $d\sigma/dy$ for photoproduction of the $a_2^+(1320)$ in $pAu$ collisions at RHIC (red) and in $pPb$ collisions at the LHC.  From Ref. \protect\cite{Klein:2019avl}.
    }
\label{fig:atwo}
\end{figure}

UPC photoproduction of the $a_2^+(1320)$ has been considered for proton targets, using parameterizations of $\sigma(\gamma p\rightarrow a_2^+ (1320) p$ \cite{Klein:2019avl}.  The photoproduction cross section is smaller than the $\rho^0$ cross section, but still large enough to be easily visible in $pA$ UPCs.  These reactions occur via Reggeon exchange, with the bulk of the cross section near threshold.  So, as can be seen in Fig. \ref{fig:atwo},  production is peaked near rapidity 2.5 at RHIC, and near 7 at the LHC.  This is outside of the acceptance of most current detectors.  The STAR forward upgrade  \cite{Sun:2023ras} covers the appropriate rapidity range., so might be used to probe $a_2^+ (1320)$ production if an appropriate trigger is available.  UPC photoproduction in $AA$ collisions should have similar characteristics, although with a shift toward midrapidity because of the lower target beam energy.   

The $a_2^\pm(1320)$ is an example of a relatively standard `standard candle' $q\overline q$ state, but other mesons can be produced, including hadronic `exotica' like 4-quark states and other hybrids.  One example is the $Z_c^+ (4430)$, where the cross section is very roughly 1/400 of that of the $a_2(1320)^+$  \cite{Klein:2019avl}.   Because of the high mass, production is somewhat shifted toward mid-rapidity.

\section{Multiple vector meson production and interference}

The cross section for multiple vector meson production can be calculated in impact-parameter space: \cite{Baltz:2002pp,Klein:1999qj}:
\begin{equation}
\sigma(V_1,V_2) = \int d^2b P(V_1,b) P(V_2,b)
\end{equation}
where $P(V_1,b)$ and $P(V_2,b)$ are the probabilities for producing the two vector mesons at impact parameter $b$.  It is easy to incorporate trigger conditions (such as single of mutual Coulomb dissociation) by adding a third probability, accounting for the probability of additional photon exchange(s), leading to the desired excitation(s).    Similar techniques can be used to calculate the kinematic distributions, such as the joint rapidity distribution \cite{Klusek-Gawenda:2013dka,Azevedo:2023vsz}.  

Even for a single vector meson, interesting interference phenomena are observable, due to the destructive interference between photoproduction on the two ions \cite{Klein:1999gv,STAR:2008llz,STAR:2022wfe}.   This interference reduces production at small $p_T$ and a given rapidity $y$:
\begin{equation}
\sigma(y,p_T) = \big|A_1(y,p_T)-A_2(y,p_T)e^{i(\vec{p}_T\cdot\vec{b})} \big|^2
\label{eq:interfere}
\end{equation}
where $A_1(y,p_T)$ and $A_2(y,p_T)$ are the amplitudes for production on the two nuclei.  The exponential is the propagator between the two targets.  The minus sign is because vector mesons have negative parity, and swapping from production on one nucleus to production on the other is equivalent to a parity inversion.  At mid-rapidity, symmetry requires $A_1(0,p_T)=A_2(0,p_T)$ and the interference is maximal; as $p_T\rightarrow 0$, production disappears \cite{Klein:2002gc}. 

Another aspect of this interference comes from the photon polarization.  The photons are linearly polarized along $\vec{b}$.  Since the plane formed by the decay products is preferentially oriented with respect to the electric field vector ({\it i. e.} to $\vec{b}$), this interference introduces an azimuthal modulation to the meson $p_T$ distribution \cite{Li:2019yzy,STAR:2022wfe}.   

\subsection{Non-identical mesons}

\begin{figure}[!tbp]
    \centering
    \includegraphics[width=0.85\textwidth]{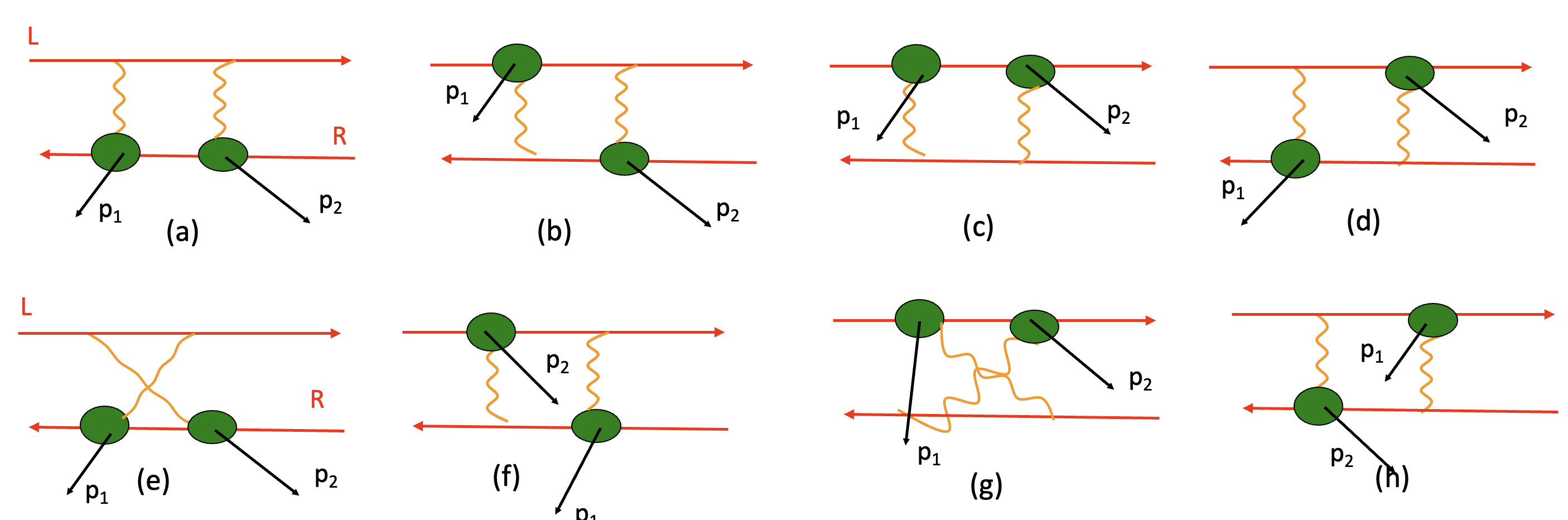} 
    \caption{(top) The four diagrams that contribute to the photoproduction of two non-identical mesons. (bottom) The four additional diagrams that contribute for the production of two identical mesons.}
\label{fig:diagrams}
\end{figure}

With two vector mesons, the interference phenomenology grows richer.  We first consider two non-identical mesons ({\it e. g.}  $\rho\phi$).  Four diagrams, shown in the top row of Fig. \ref{fig:diagrams}, contribute to two--non-identical-meson production.  The joint probability (at a given $b$) for production of two mesons follow from these diagrams:
\newcommand*{\bb}{\vec{b_i}}
\newcommand*{\bbl}{\vec{b_l}}
\newcommand*{\bbions}{\vec{b_{12}}}

\newcommand*{\pp}{\vec{p_j}}
\newcommand*{\kk}{\vec{k_j}}
\newcommand*{\qq}{\vec{q_j}}

\newcommand*{\ppa}{\vec{p_1}}
\newcommand*{\kka}{\vec{k_1}}
\newcommand*{\qqa}{\vec{q_1}}

\newcommand*{\ppb}{\vec{p_2}}
\newcommand*{\kkb}{\vec{k_2}}
\newcommand*{\qqb}{\vec{q_2}}
\newcommand*{\expikba}{e^{i(\kka\cdot\vec{b})}}
\newcommand*{\expikbb}{e^{i(\kkb\cdot\vec{b})}}

\begin{equation}
\begin{split}
P(b) = \bigg| 
& A_L(\ppa,b)A_L(\ppb,b)
 - A_L(\ppa,b)A_R(\ppb,b) \expikbb \\
& - A_R(\ppa,b)A_L(\ppb,b) \expikba
+ A_R(\ppa,b)A_R(\ppb,b) \expikbb \expikba 
\bigg|^2
\label{eq:allnonidentical}
\end{split}
\end{equation}
where $A_L$ and $A_R$ are the amplitudes for production on the left-going and right-going nuclei respectively, $\vec{p}_1$ and $\vec{p}_2$ are the vector meson momenta and $\vec{k}_1$ and $\vec{k}_2$ are the two momentum transfers from the nucleus.  Equation \ref{eq:allnonidentical} can be factorized into the product of two independent interference terms for two single mesons, and much of the phenomenology follows from that factorization: the cross section goes to zero when either one of the mesons is at mid-rapidity and has small $p_T$.  

\subsection{Identical mesons}

For a pair of identical mesons, there are four additional diagrams to consider, shown on the bottom row of Fig. \ref{fig:diagrams}, since there are two routes to create each meson - from photon 1 or photon 2.  This leads to a more complex phenomenology.  One simple but interesting case involves the emission of light vector mesons at relatively forward rapidities.   Light vector mesons are notable because $\sigma(\gamma A\rightarrow VA)$ increases only slowly with increasing photon energy, while the photon spectrum scales roughly as $1/k$.  So, photoproduction is largest near threshold, corresponding to production at forward rapidity.   In this region, the two leftmost diagrams  (or, for negative rapidity, the third column of Fig. \ref{fig:diagrams}) are the dominant contribution, and 
\begin{equation}
\begin{split}
P(b) = \bigg| \Sigma_i\Sigma_j
& A_L(\ppa,b)A_L(\ppb,b) \exp(i(\vec{k}_1\cdot\vec{b}_i+\vec{k}_2\cdot\vec{b}_j)) \\
& + A_L(\ppb,b)A_L(\ppb,a) \exp(i(\vec{k}_2\cdot\vec{b}_i+\vec{k}_1\cdot\vec{b}_j))
\bigg|^2.
\end{split}
\label{eq:forward}
\end{equation}
The sums over $i$ and $j$  run over the target nucleons at positions $\vec{b}_i$.  They exhibit the conditions for coherent emission from the nucleus.  Essentially, these boil down to $|\vec{k}_1|,|\vec{k}_2| < \hbar/R_A$.

Within this low $p_T$ constraint, Eq. \ref{eq:forward}  of the two mesons leads to two identical terms for the amplitude.   If we neglect shadowing or other geometric effects then the production amplitudes are all identical.  This assumption is good for heavier mesons like the $J/\psi$, but less accurate for lighter mesons where shadowing is large and a Glauber calculation is needed to find the total cross section.  But, it simplifies the analysis, while retaining the essential physics, so we will use that assumption here.   With this, at low $p_T$, the sums over $i$ and $j$ can be replaced with $N$ (the number of nucleons in the target nucleus) times the individual amplitudes.  If both vector mesons are produced at similar rapidities, so the amplitudes are similar, the cross section is
\begin{equation}
P(b) = |2N^2   A_L(\ppa,b)^2|^2 = 4 N^4 A_L(\ppa,b)^4.
\label{eq:twoforward}
\end{equation}
The cross section is double what it would be without the additional identical-particle diagrams.  This is an example of superradiance.

The coherence conditions are  tighter than they are for a single meson, because the exponentials in Eq. \ref{eq:forward} sum two products of momentum and impact parameter.  So, the coherence condition becomes $|\vec{k}_1|,|\vec{k}_2| < \hbar/2R_A$. 

It is easy to generalize this to the emission of more than two mesons.  For the emission of $M$ mesons, there are $M!$ independent diagrams.  So, the amplitude for emission, subject to all of the mesons having sufficiently low $p_T$ is
\begin{equation}
P_M(b) = M! N^{M*M} A_L(\ppa,b)^{M*M} = M! P_1(b)^M
\label{eq:laser}
\end{equation}
The production probability is $M!$ times the probability if superradiance is neglected.  For large $M$, $M!$ increases faster than $P_1(b)^M$.  For producing $\rho$ mesons with lead beams at the LHC, $P_1(b=2R_A) \approx 0.03$ \cite{Klein:1999qj}; this might rise to $\approx 0.05$ at a future higher-energy accelerator; it would also be somewhat higher with uranium or other higher$-Z$ beams.  

Whatever $P_1(b)$ is, if enough vector mesons are produced, the system can reach the criteria to be considered a laser.  In optical terms, ``the optical gain equals the cavity loss'' \cite{lasing}.   Here, the equivalent condition is  that  $P(M+1) > P(M)$.  This requires $M$ to be quite large ($M>30$ for lead at the LHC), and it is likely that other limitations would come into play.  First, the coherence conditions tighten as $M$ rises and the number of terms in the exponential increases, to $\vec{k}< \hbar/MR_A$.  This has similarities with optical systems, where the beam divergence drops as lasing sets in.  Another limitation is energetic, since the energy carried by the electromagnetic field is finite.   Although the large$-M$ limit may be beyond current reach, it should be possible to observe superradiance from two or three vector mesons.

\subsection{Decay correlations and stimulated decays}

Correlations should also be visible in the particle decays.  One effect is straightforward: the vector mesons share the same linear polarization, so there will be correlations in decay angles.   Another possibility is more intriguing: if the two vector mesons are in the same quantum mechanical state, there is the possibility of stimulated decays, where the probability for decays to the identical final state is enhanced.  

Stimulated decays can occur for final states particles that have integer spin (i. e. bosons).  If there is already an identical particle in the same state that will result from the decay, the probability for that decay can be enhanced. Stimulated decays have previously been studied in the context of transitions between atomic states, especially the de-excitation of excited states \cite{DeMartini:1988}.   Stimulated decay of dark matter axions has been recently considered \cite{Caputo:2022keo}.  A field of photons can stimulate the decay of axions when one of the decay photons has a momentum matching that of the field.  

 They are in the same state if their momentum is close enough together that they are still in-phase when the first one decays; the field of final-state particles stimulates decay to that same final state \cite{Caputo:2018vmy}.  The possibility of stimulated meson decay inside nuclei (where final state mesons would already be present) has also been proposed \cite{Furry}.    In UPCs, stimulated decay could appear in decays like $\rho\rightarrow\pi^+\pi^-$, where the $\pi^\pm$ from multiple decays might cluster in phase space.   In contrast, for decays like $J/\psi\rightarrow e^+e^-$, the fermionic statistics should preclude  final state lepton clustering. 

\section{Conclusions}

Coherence is an interesting and subtle topic, with important consequences.  The Good-Walker paradigm relates coherent and incoherent photoproduction to the average nuclear configuration and configuration fluctuations respectively.  In Good-Walker coherence means that the target remains in the ground state.  That disagrees with some experimental observations of coherent photoproduction. An alternate approach to coherence adding amplitudes, can explain data where Good-Walker fails. The 'adding amplitudes' approach allows some new coherent UPCs channels, including coherent photoproduction of charged mesons.

UPC production of multiple vector mesons introduces several new quantum mechanical observables.  When multiple mesons are produced in the forward direction, then superradiant emission may occur, increasing the probability of multiple mesons.    It may even be possible to observe the stimulated decay of vector mesons. 

\section*{Acknowledgments}
 
 I thank Minjung Kim for useful comments on this manuscript.  This work is supported in part by the U.S. Department of Energy, Office of Science, Office of Nuclear Physics, under contract numbers DE-AC02-05CH11231.

\end{document}